\documentclass[twocolumn,prl,aps]{revtex4}

 \usepackage[dvips]{graphicx}
 \usepackage[dvips]{graphics}

\usepackage{ulem}

\usepackage{array}
\usepackage{multirow}
\usepackage{hhline}
\usepackage{bigstrut}
\usepackage{slashbox}

\newlength{\bxwidth}\bxwidth=0.8\textwidth

\begin{document}
\title{Doping Dependence of Bilayer Resonant Spin Excitations in $\bf (Y,Ca)Ba_2Cu_3O_{6+x}$}
\author{S. ~Pailh\`es$^{1,2}$, C. Ulrich$^3$, B. Fauqu\'e$^1$,
V. Hinkov$^3$, Y. ~Sidis$^1$,  A. ~Ivanov$^4$, C.T. Lin$^3$,
 B. Keimer$^{3}$ and P.~Bourges$^{1\ast}$}

\affiliation{
$^1$ Laboratoire L\'eon Brillouin, CEA-CNRS, CE-Saclay, 91191 Gif sur Yvette, France.\\
$^2$ Laboratory for Neutron Scattering, ETH Zurich and PSI Villigen, CH-5232 Villigen PSI, Switzerland.\\
$^3$ Max-Planck-Institut f\"ur Festk\"orperforschung, 70569 Stuttgart, Germany.\\
$^4$ Institut Laue Langevin, 156X, 38042 Grenoble cedex 9, France.
}


\begin{abstract}
Resonant magnetic modes with odd and even symmetries were studied
by inelastic neutron scattering experiments in the bilayer
high-$T_c$ superconductor $\rm Y_{1-x}Ca_{x}Ba_2Cu_3O_{6+y}$ over
a wide doping range. The threshold of the spin excitation
continuum in the superconducting state, deduced from the energies
and spectral weights of both modes, is compared with the
superconducting $d$-wave gap, measured on the same samples by
electronic Raman scattering in the $B_{1g}$ symmetry. Above a
critical doping level of $\delta \simeq 0.19$, both mode energies
and the continuum threshold coincide. We find a simple scaling
relationship between the characteristic energies and spectral
weights of both modes, which indicates that the resonant modes are
bound states in the superconducting energy gap, as predicted by
the spin-exciton model of the resonant mode.
\end{abstract}

\maketitle


In high-$T_c$ copper oxides superconductors, inelastic neutron
scattering (INS) experiments have shown that the superconducting
(SC) phase exhibits an unusual spin triplet excitation, the
so-called magnetic resonance mode \cite{Sidis_review}. This
excitation has been observed in several families of copper oxides
with SC critical temperatures $\rm T_c \sim$90 K: ${\rm
Tl_2Ba_2CuO_{6+x }}$, $\rm Bi_2Sr_2CaCu_2O_{8+x}$, and $\rm
YBa_2Cu_3O_{6+x}$(YBCO$_{6+x}$). Two kinds of theoretical models
attempt to describe this mode. Within an itinerant-electron
picture, it is described as a {\it spin-exciton}, that is, it
corresponds to a spin-triplet bound states stabilized by the
electron-electron interaction in the plane below the
electron-hole ({\it e-h}) spin-flip continuum, which is gapped in
the SC state \cite{Millis96,eshrig}. As the mode is an
intrinsic feature of the SC state, it disappears above $T_c$. In
other approaches \cite{prexist,stripes1,stripes2}, the collective
spin excitations pre-exist in the normal state, where they are
damped by scattering from charge excitations.
In the SC state, the damping is suppressed, and the mode sharpens
if its energy is below the threshold of the gapped {\it e-h}
continuum. The charge carriers interacting with the spin
excitations can be either uniformly distributed \cite{prexist}
or segregated into quasi-one-dimensional {\it stripes} that
separate nearly insulating antiferromagnetic domains
\cite{stripes1,stripes2}. Although the quantum numbers
characterizing the mode are identical in both of these competing
approaches, the underlying physical pictures are quite different.
It is hence important to establish criteria that allow
quantitative experimental tests of these models.

The bilayer structure of YBCO offers an interesting opportunity in
this regard. While most calculations have been carried out for a
single $\rm CuO_2$ plane per unit cell
\cite{stripes1,prexist}, both theoretical approaches
predict two distinct magnetic resonant modes in a bilayer system.
They are characterized by even ($e$) or odd ($o$) symmetries,
respectively, with respect to the exchange interaction between the
 two layers in a bilayer unit. Recent INS measurements have shown that the bilayer
system YBCO indeed exhibits two magnetic resonant modes, both of
which are apparent as a strong enhancement of the magnetic
intensity at the planar antiferromagnetic wave vector ${\bf
q_{AF}}=(\pi,\pi)$ in the SC state
\cite{PRL_Pailhes04,PRL_Pailhes03}. Their intensities display an
order-parameter-like temperature dependence and vanish at $\rm
T_c$, without any significant shift of their characteristic
energies. A crucial difference between both theoretical approaches
concerns the energy-integrated spectral weight (SW) of the
magnetic modes. In {\it spin exciton} models, this quantity is a
function of the binding energy of the modes\cite{Millis96,eshrig}(that is, the
difference of the mode energy and the {\it e-h} continuum
threshold), whereas in models based on {\it pre-existing} spin
excitations it
is not obviously related the location of the mode with respect to
the {\it e-h} continuum.

Here we report an INS study of the odd and even resonant spin
excitations at ${\bf q_{AF}}$ spanning a wide doping range from
the underdoped to the overdoped regimes in YBCO. The threshold for
{\it e-h} excitations, 2$\rm \Delta_{max}$, was measured by
electronic Raman scattering (ERS) in the $\rm B_{1g}$ channel
\cite{Gallais} on the same samples used in the neutron
experiments. This eliminates systematic errors invariably
associated with comparisons of measurements on samples from
different origins. We found that the resonant spin excitations are
always located below 2$\rm \Delta_{max}$. Assuming that the spin
excitations are bound states in the energy gap of the {\it e-h}
continuum, its threshold at ${\bf q_{AF}}$, $\omega_c$, is
estimated using the measured energies, $\rm E_r^{o,e}$, and
spectral weights, $\rm W_r^{o,e}$, of the odd and even resonance
peaks. The threshold energy $\omega_c$ estimated in this way is
found to be below the independently determined gap 2$\rm
\Delta_{max}$, so that the analysis is self-consistent. We also
observed a systematic scaling relation between $\rm W_r^{o,e}$ and
the reduced binding energy ($\omega_c$-$\rm
E_r^{o,e}$)/$\omega_c$. As such a relation is only expected for
collective bound states of the {\it e-h} continuum, our results
support the spin-exciton description of the resonant mode.
Further, our study reveals that 2$\rm \Delta_{max}$ and the total
SW of the resonant spin excitations at ${\bf q_{AF}}$ drop
precipitously in the overdoped regime close to the hole doping
level $\delta \simeq$0.19. Previous work identified of this doping 
level as the end point of the
pseudo-gap phase \cite{Tallon_99,Tallon,Naquib}.

In addition to previous INS experiments on a slightly overdoped
sample (OD85, Y$_{0.9}$Ca$_{0.1}$Ba$_2$Cu$_3$O$_{7}$, Ref.
\cite{PRL_Pailhes03}) and a slightly underdoped sample (UD89),
YBa$_2$Cu$_3$O$_{6.85}$, Ref. \cite{PRL_Pailhes04}), INS studies
of bilayer excitations were performed on two YBCO samples. The
first sample is overdoped (OD75),
Y$_{0.85}$Ca$_{0.15}$Ba$_2$Cu$_3$O$_{7}$ with $T_c=75$ K\cite{JSSC_Carron89,Tallon95}.
Following our previous work with sample OD85, about 50 single
crystals, obtained by a top-seeded solution growth method
\cite{JCG_Lin02}, were co-aligned on aluminum plates and fixed
with glue or Al screws. The second sample (UD63),
YBa$_2$Cu$_3$O$_{6.6}$ with $T^{onset}_c=63$ K, is underdoped and
consists of 180 detwinned squared-shape single crystals
\cite{Nature_Hinkov04}. The mosaic spread of these
crystals arrays exceeds 1.4$^\circ$. INS measurements have been
performed on the thermal triple axis spectrometer IN8 at the Institute Laue 
Langevin (Grenoble). The experimental setup and scattering plane 
are the same as Ref. \cite{PRL_Pailhes04} except that we utilized
a PG(002) monochromator. A fixed final neutron energy 
of 35 meV has been used yielding a typical energy resolution of 7-8 meV.

The spin susceptibility of a bilayer system reads: 
\begin{eqnarray}
\chi({\bf Q},\omega) =  \Big[ \sin^2 (\pi z L) \chi_{o}({\bf
q},\omega) + \cos^2(\pi z L)\chi_{e}({\bf q},\omega) \Big]
\label{eq-bilayer}\end{eqnarray} where ${\bf{Q}}=(H,K,L)$ is the
full wave vector and  ${\bf{q}}=(H,K)$ is the planar wave vector
in  CuO$_2$ plane. z=0.28 stands for the reduced distance between
the planes of the bilayer. The L-component of the wave vector can
thus be used to select either the odd or the even channel
\cite{PRL_Pailhes04}: the magnetic structure factor is maximum for
L$\sim$ 5.2 in the odd channel and L$\sim$ for 3.4 or 7 in the
even. Following the standard terminology
\cite{Fong,PRL_Pailhes03,PRL_Pailhes04,PRB_Pailhes05}, we define
the magnetic resonant mode as the enhancement of the imaginary
part of the magnetic susceptibility, $\Delta Im \chi({\bf
q_{AF}},\omega$) in the SC state, which is derived from the
difference between measurements in the SC state at 10 K and the
normal state above $\rm T_c$.
The imaginary part of the dynamical magnetic
susceptibility is also calibrated in absolute units ($\rm
\mu_B^2.eV^{-1}/f.u.$) following a standard procedure with a
reference phonon at 42.5 meV \cite{Fong,PRB_Pailhes05}.

\begin{figure}[t]
\includegraphics[width=7cm,angle=270]{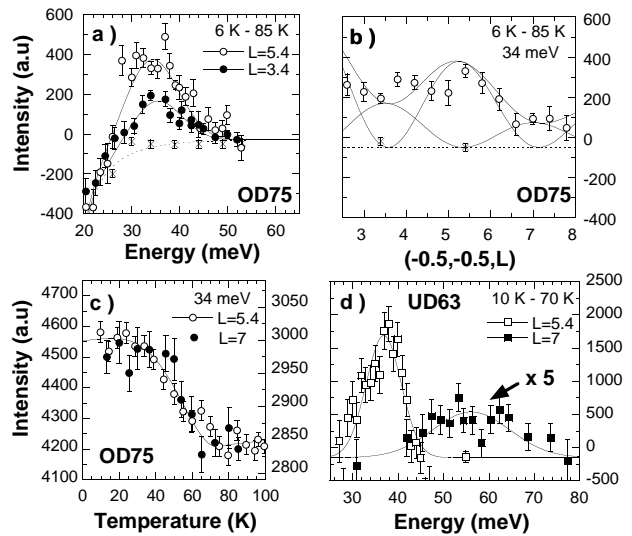}
\caption { (color online) a) Differences between INS energy scans
measured at the antiferromagnetic wave vector below and below $\rm
T_c$ in Y$_{0.85}$Ca$_{0.15}$Ba$_2$Cu$_3$O$_{7}$ (OD75). Open and
closed symbols correspond to the odd and even channels,
respectively. The data have been corrected by the anisotropic Cu 
magnetic form factor. The stars indicate the measured reference level of
the magnetic scattering determined by constant-energy scans. 
b) L-dependence at E=34 meV for OD75. The lines describe the intensity
of each mode following Eq. \ref{eq-bilayer} multiplied by the Cu
magnetic form factor (upper line). c) Temperature dependence of
the neutron intensity measured at characteristic energies of the
even and odd modes for OD75. d) Same as a) but for sample
YBCO$_{6.6}$ (UD63).} {\label{fig1}}
\end{figure}

Fig. \ref{fig1}.a provides evidence of the existence of odd
and even resonant magnetic peaks in the INS spectra of the highly overdoped sample
OD75. These results differ significantly from previous reports at
lower doping levels \cite{PRL_Pailhes03,PRL_Pailhes04}, because
the energies of both mode are very similar: E$_r^{o}$=34meV and
E$_r^{e}$=35meV. Further, both modes also display similar
amplitudes. In order to determine the amplitudes $\rm I^{o,e}$ of
both modes, the L-dependence of the magnetic intensity at 34 meV
(Fig. \ref{fig1}.b) was fitted with Eq. \ref{eq-bilayer}. This
yields a ratio ${\rm I^{e}/I^{o}\simeq 0.4}$, compared to 1/3 for
sample OD85 \cite{PRL_Pailhes03} and 1/6 for YBCO$_{6.85}$ 
\cite{PRL_Pailhes04}. However, as for lower doping levels
\cite{PRL_Pailhes03,PRL_Pailhes04}, the temperature dependence of
the intensities of both modes exhibits a marked change at the SC
temperature (Fig. \ref{fig1}.c), indicating that the phenomena
have a similar origin. In the strongly underdoped sample UD63 (Fig.
\ref{fig1}.d), the two mode energies and their intensities are
very different: E$_r^{o}$=37 meV, E$_r^{e}$=55 meV, and ${\rm
I^{e}/I^{o}\simeq 1/20}$. As already observed in the OD85 sample
\cite{PRL_Pailhes03}, the odd mode exhibits a broader energy line
shape in OD75, $\Delta$E$\sim$ 12 meV, than the resolution-limited
energy profile observed in underdoped samples
\cite{Nature_Hinkov04}. The energy width of the even mode,
$\Delta$E$\sim$ 15 meV, is always broader than the resolution,
for all doping level.

\begin{figure}[t]
\centerline{\includegraphics[angle=270,width=8 cm]{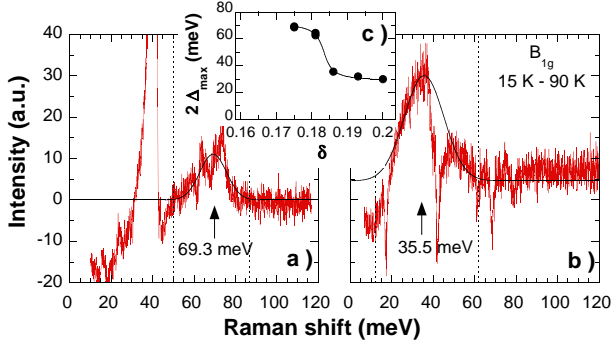}}
\caption{(color online) Electronic Raman spectra in the $\rm
B_{1g} channel$. The difference between spectra measured at 15 K
in the SC state and at 90 K in the normal state is
displayed. In addition to sharp features due to phonon
renormalization, the difference indicates a broad peak due to the
enhancement of the electronic response in the SC state. The
electronic peaks were fitted with Gaussian profiles. The inset shows
the peak energy of the 5 samples we have studied (see
Fig.~\ref{doping_dep} caption for the determination of the doping
levels). } \label{ERS_spectrum}
\end{figure}

In order to assess whether or not these resonant modes should be
regarded as bound states in the electronic continuum, one has to
determine the amplitude of the SC gap. The good surface quality of
the single crystals removed from the crystals array OD85 and OD75
enables ERS experiments. The ERS technique
probes electronic excitations in selected areas of the Brillouin
zone. In the $\rm B_{1g}$ channel, the antinodal areas around ($
\pm \pi, 0$) and (0,$ \pm \pi $) are probed, and a peak appears in
the electronic spectrum in the SC state at twice the energy of the
maximum $d$-wave SC gap, 2$\rm \Delta_{max}$\cite{Gallais}.
 These measurements were carried out with
a triple-grating spectrometer in quasi-back-scattering geometry.
The crystals were mounted on the cold finger of an He circulation
cryostat. The 514 nm excitation line of a Ar$^+$/Kr$^+$ mixed-gas
ion laser was used.

Figure \ref{ERS_spectrum} shows the $\rm B_{1g}$ peaks in the ERS
spectra of single crystals extracted from samples OD85 and OD75. The
$\rm B_{1g}$ peak is observed at 69.3 meV in the first sample and
shifts to 35.5 meV for the second, whereas $\rm T_c$ is only
reduced from 85.5 K to 75 K. The same results were obtained on
other crystals selected randomly from the OD85 and OD75 arrays\cite{pailhes_futur}.
Further experiments on samples with different $T_c$ also confirm
that the characteristic energy of the $\rm B_{1g}$ peak drops
steeply by almost a factor two with a minor reduction of $\rm
T_c$. This phenomenon was previously reported in
Refs.~\cite{Chen_PRB93,Bok_AP99}, and it is now confirmed directly
on the same samples used for the INS experiments. Systematic
errors associated with experiments on samples of different origins
could thus be avoided. As shown in the inset of
Fig.~\ref{ERS_spectrum}, the drastic decrease of the $\rm B_{1g}$
peak occurs close to a critical hole doping level
$\delta_c\simeq$ 0.19.

\begin{figure}[t]
\centerline{\includegraphics[angle=270,width=8cm]{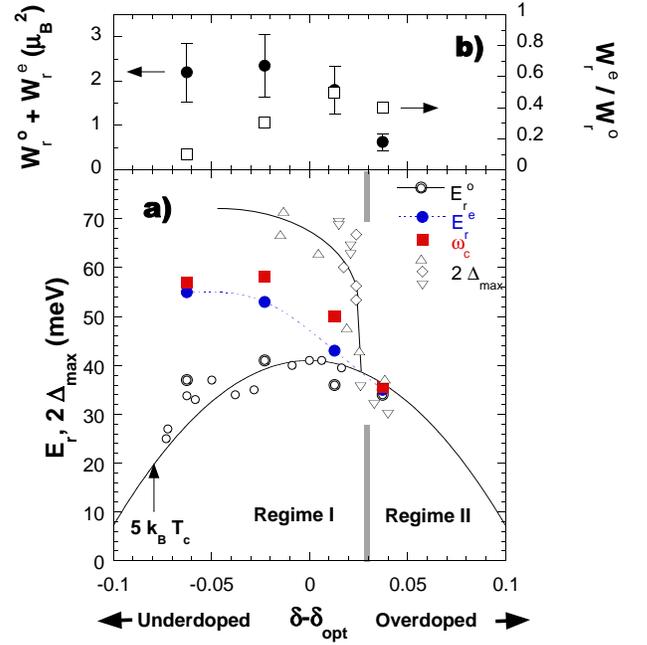}} 
\caption{(color online) Doping
dependence of: a) $\rm E_r^o$, $\rm E_r^e$, $\omega_c$ and 2$\rm
\Delta_ {max}$(ERS). Other ERS data were obtained from \cite{Bok_AP99}
($\bigtriangleup$), in \cite{Chen_PRB93} ($\diamond$), in the
present study ($\bigtriangledown$).  b) Ratio and sum of the
energy-integrated SW. The hole doping is estimated from the
phenomenological relation $\rm
T_c(\delta)=T_c^{opt}(1-82.6(\delta-\delta_{opt})^2)$
\cite{Tallon95}, $\delta_{opt}$ is generally estimated to be
0.16 \cite{Tallon95}. $\rm T_c^{opt}$ is 93K for pure YBCO. 
Ca substitution reduces T$_c$ recording to $\rm T_c^{opt}= 90.6 - 39.2 x_{Ca}$ 
obtained for our single crystals. } \label{doping_dep}
\end{figure}

In Fig.~\ref{doping_dep}.a, we report the hole doping dependence (i) 
of the of 2$\rm \Delta_{max}$ determined by ERS from the
present study and Refs.~\cite{Chen_PRB93,Bok_AP99}, and (ii) the
odd and even resonant energies obtained from INS data. For all
samples, the even resonance appears at higher energy and with less
intensity than the odd resonance. While $\rm E_r^o$ evolves as a
function of hole doping as $\sim$ 5 $\rm k_B T_c$
(Fig.~\ref{doping_dep}.a), $\rm E_r^e$ does not scale with $\rm
T_c$: it remains almost constant in the underdoped regime and
begins to decrease upon entering the overdoped regime. With
increasing hole doping, the distance in energy between the
magnetic modes decreases from 17 meV in UD63 to 1 meV in OD75. Owing
to the differences between the energy lineshapes of the odd and
even modes, the most meaningful comparison is based on their
energy-integrated spectral weights: $W_r^{o,e}$=$\int_0^\infty$
d$\omega \Delta Im \chi_{o,e}({\bf q},\omega)$. The ratio and the
sum of the energy-integrated SWs of both modes are shown in
Fig.~\ref{doping_dep}.b. Obviously, $W_r^{e}/W_r^{o}$ increases
with increasing doping.

The dramatic change of the SC gap measured by ERS
(Fig.~\ref{doping_dep}.a) separates two distinct regimes on both
sides of $\delta_c$: in regime I, encompassing the underdoped
samples up to OD85, one obtains the following hierarchy: $\rm 5 k_B
T_c \sim \rm E_r^o < E_r^e < \rm 2 \Delta_{max}$. In regime II
(OD75), all energies collapse to $\rm 5 k_B T_c \sim E_r^o \simeq
E_r^e \simeq \rm 2 \Delta_{max}$. The coincidence of the resonant
mode energy and the SC energy gap in the overdoped range was
suggested by prior tunneling experiments \cite{zasadzinski}. As the information 
about the mode energy obtained from tunneling
is quite indirect, the interpretation of these data has remained
controversial. The total SW of the resonant modes,
$W_r^{o}+W_r^{e}$, declines precipitously around a similar doping
level (Fig.~\ref{doping_dep}.b). It is interesting to note that
the doping level $\delta_c \simeq$ 0.19 separating these two
regimes coincides with the hole concentration where previous work
had uncovered rapid doping-induced modifications of the physical
properties such as resistivity, specific heat, Knight shift, and
muon spin relaxation rate \cite{Tallon_99,Tallon,Naquib}.
According to \cite{SachdevRev}, this point in the phase diagram is a
quantum critical point associated with the doping-induced
disappearance of a putative order parameter characterizing the
pseudo-gap phase. Therefore, it is tempting to
relate the anomalous hole doping dependencies shown in
Fig.~\ref{doping_dep}.a to the pseudo-gap phenomenon.

In Fig.~\ref{doping_dep}.a, one also notices that the resonance
energies are systematically lower than twice the SC gap. This
indicates that the resonant modes are below the continuum
threshold at ${\bf q_{AF}}$, $\omega_c$, which can be estimated as
1.8$\rm \Delta_{max}$ using the typical Fermi surface topology
observed in cuprates \cite{kordyuk}. As discussed in Refs.
\cite{PRL_Pailhes03,Millis96}, another estimate of $\omega_c$ can
be made within the spin-exciton model where $W_r^{o,e}\propto
(\omega_c - E_r^{o,e})/\omega_c$. 
$\omega_c$ is simply deduced by the ratio
of the SWs of the odd and even modes
\cite{PRL_Pailhes03,PRL_Pailhes04}. For the four samples, $\omega_c$ extracted
in this way from INS (Fig.~\ref{doping_dep}.a) agrees with the value
deduced from ERS data.

\begin{figure}[t]
\centerline{\includegraphics[angle=270,width=7 cm]{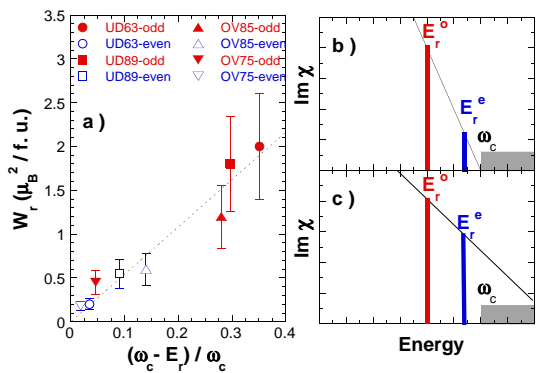}}
\caption{(color online)
 a) Resonant modes SW {\it vs} the reduced binding energy.
Sketch of two approaches for the resonant modes:
 b) bound states of the electronic continuum. c) pre-existing modes whose
intensities are enhanced by the disappearance of the fermionic
damping below $\omega_c$.} \label{fig4}
\end{figure}

We now discuss these observations in the light of the various
models attempting to describe the resonant modes
\cite{Millis96,eshrig,prexist,stripes1,stripes2}. As
sketched in Fig.~\ref{fig4}, one can envisage two situations:
either the modes are bound states in the energy gap of the
electronic continuum (Fig.~\ref{fig4}.b), as assumed in the {\it
spin-exciton} model \cite{Millis96,eshrig}, or they
already exist above $T_c$ \cite{prexist,stripes1,stripes2} and
their SWs are only enhanced in the SC state (Fig.~\ref{fig4}.c). 
In the former case, the SWs of the modes are proportional to the binding energies,
while in the latter case there is no specific relation between the
SW and the distance between the mode energy and $\omega_c$. The SWs of the modes, 
$W_r^{o,e}$, are
presented in Fig.~\ref{fig4}.a as a function of the reduced
binding energy, i.e. $(\omega_c-E_r^{o,e})/\omega_c$. Clearly,
these two quantities are proportional to each other with a constant
scaling factor over a wide doping level. This is the expected behavior 
for collective bound states in the SC energy gap where the scaling 
factor is mostly controlled by doping independent microscopic 
parameters.
In the pre-existing mode picture, on the other hand, the SW ratio
$W_r^{e}/W_r^{o}$ would be also determined by the spin dynamics in
the normal state. To be
sure, well-defined magnetic modes are observed in the underdoped
samples above $T_c$ \cite{Fong}, but we recently found that there
are a number of qualitative differences between the excitation
spectra in the SC and normal states at energies comparable to the
SC gap \cite{hinkov05}. In overdoped samples, normal-state
excitations appear to be heavily overdamped and have thus far not
been clearly identified. It is therefore reasonable to consider
the excitations in the SC state separately, as we have done.

In summary, odd and even symmetry resonant magnetic excitations
have been observed in bilayer YBCO over a wide doping range. A
detailed analysis of the modes energies and spectral weights show
that the resonant mode in the high-$T_c$ cuprates arises from bound
states in the superconducting energy gap. Our results have no
natural explanation in models where the resonant excitations are
associated with excitations pre-existing in the normal state.

\end{document}